\documentclass[twocolumn,showpacs,preprintnumbers,amsmath,amssymb]{revtex4}

\usepackage{graphicx}
\usepackage{dcolumn}
\usepackage{bm}

\begin{document}


\title{EGRET Excess of diffuse Galactic Gamma
Rays\\ interpreted as a Signal of Dark Matter Annihilation}

\author{W. de Boer}
\email{Wim.de.Boer@cern.ch}
\author{C. Sander}
\author{V. Zhukov}
 \affiliation{Institut f\"ur Experimentelle Kernphysik
Universit\"at Karlsruhe (TH), P.O. Box 6980, 76128 Karlsruhe,
Germany}

\author{A.V. Gladyshev, D.I.  Kazakov}%
\affiliation{Bogoliubov Laboratory of Theoretical Physics, Joint
Institute for Nuclear Research, 141 980 Dubna, Moscow Region,
Russia}

\date{\today}

\pacs{95.35.+d, 12.60.Jv, 98.70.Rz, 98.80.Cq}
\maketitle

Elsaesser and Mannheim~\cite{elsaesser} fit a contribution of Dark
Matter Annihilation (DMA) to the {\it extragalactic} contribution of
the galactic diffuse gamma ray flux, as deduced from the EGRET data
by Strong, Moskalenko and Reimer~\cite{optimized}.

\begin{figure}[h]
\begin{center}
 \includegraphics [width=0.4\textwidth,clip]{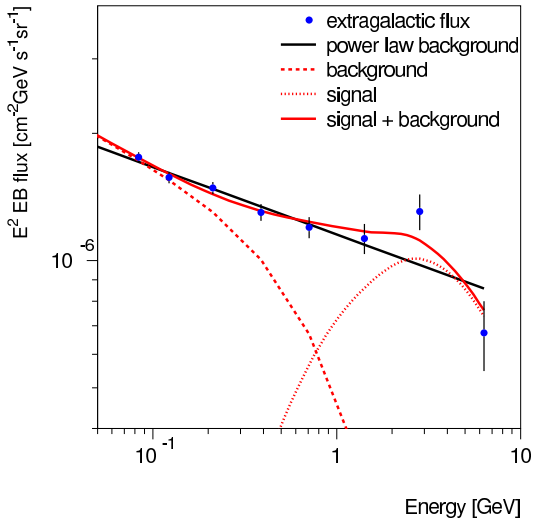}
 \caption{The extragalactic flux of diffuse gamma rays compared with a
simple power low (thick (black) line) and the sum of a steeply
falling background plus a WIMP annihilation signal (grey (red) line
with the components as dashed lines).} \label{f1}
\end{center}
\end{figure}
 They find a WIMP
mass of 515$^{+110}_{-75}$ GeV and quote a systematic error of 30\%.
However, they do not include large systematic 
uncertainties from the fact that    the determination of the extragalactic
flux (EGF) requires a model for the subtraction of the Galactic flux from the
data. The data used were  obtained with a model without
Galactic DMA, so one expects additional uncertainty in the region where  DMA
contributes. This is demonstrated in Fig. \ref{f1}, where the EGF is
obtained by subtracting  the Galactic contribution
\textit{including} the contribution from DMA~\cite{sander}. The
latter was determined from the excess of the EGRET data above the background from nuclear interactions in all sky
directions, which allows to obtain a parametrization of the halo
profile. This halo profile was shown to describe the peculiar shape
of the rotation curve of our Galaxy~\cite{deboer}, thus proving
that the Galactic excess of EGRET data traces the DM.  With this DM halo
profile the total Galactic flux including DMA can be calculated in
all directions and subtracted from the EGRET data using the
pioneering method of Sreekumar et al.~\cite{sreekumar}. This
procedure was repeated for 8 different energy bins and the results
are plotted in Fig.\ref{f1}. As expected, the high energy tail differs considerably 
 from Ref. \cite{optimized}  and can be either fitted 
with a simple power law, which yields a $\chi^2/d.o.f$ of 10.9/6 or
a probability of 9\%, or by a double power law  plus a
contribution from DMA with a WIMP mass of 50 GeV, which yields a
$\chi^2/d.o.f$ of 4.7/4 or a probability of 31\%. 
For the latter fit the shape of DMA was taken from Ref. \cite{deboer}
and the shape of the remaining contribution of the EGF could
be fitted with a double power law, typical of many point sources.
 Both probabilities
are acceptable, so there is no evidence for a signal of DMA in the
\textit{extragalactic} flux, but on the other hand a WIMP mass of 50
GeV is certainly acceptable and compatible with the excess in the
\textit{Galactic} data \cite{deboer}, but incompatible with the value given in
Ref.~\cite{elsaesser}.


\begin{thebibliography}{99}
\bibitem{elsaesser}
D. Elsaesser and K. Mannheim, ``Supersymmetric dark matter and the
extragalactic gamma ray background,'' Phys.\ Rev.\ Lett.\  {\bf 94}
(2005) 171302; astro-ph/0405235.
\bibitem{optimized}
A.W. Strong, I.V. Moskalenko and O. Reimer, ``Diffuse Galactic
continuum gamma rays. A model compatible with EGRET data and
cosmic-ray measurements,'' Astrophys. J. \textbf{613} (2004) 962;
astro-ph/0406254.
\bibitem{sander} C. Sander, Ph.D Thesis, Univ. Karlsruhe (2005)
\bibitem{deboer}
W.~de Boer et al.,
``Excess of EGRET diffuse Galactic gamma ray data as 
Tracer of Dark Matter'', A\&A, 444 (2005) 51, astro-ph/0508617;\\
W.~de Boer,  New Astronomy Reviews, \textbf{49} (2005) 213;
hep-ph/0408166.
\bibitem{sreekumar} P. Sreekumar et al., Astrophys.J. \textbf{494} (1998)
523.
\end{thebibliography}

\end{document}